\documentclass{article}

\usepackage{booktabs}
\usepackage{graphics}
\usepackage{graphicx}
\usepackage{listings}
\usepackage{nameref}
\usepackage{algorithm}
\usepackage{algpseudocode}
\algrenewcommand\algorithmicrequire{\textbf{Input:}}
\algrenewcommand\algorithmicensure{\textbf{Output:}}
\usepackage{bm}
\usepackage{amsmath}
\usepackage{amssymb}
\usepackage{caption} 
\usepackage{multirow}
\usepackage{color}
\usepackage[table]{xcolor}

\usepackage{arxiv}

\usepackage[utf8]{inputenc} 
\usepackage[T1]{fontenc}    
\usepackage{hyperref}       
\usepackage{url}            
\usepackage{booktabs}       
\usepackage{amsfonts}       
\usepackage{nicefrac}       
\usepackage{microtype}      
\usepackage{lipsum}
\usepackage{graphicx}
\graphicspath{ {./images/} }

\title{PhaVIP: Phage VIrion Protein classification based on chaos game representation and Vision Transformer}

\author{
 Jiayu Shang \\
  Dept. of Electrical Engineering\\
  City University of Hong Kong\\
  Kowloon, Hong Kong SAR, China\\
  \texttt{jyshang2-c@my.cityu.edu.hk} \\
  \And
 Cheng Peng \\
  Dept. of Electrical Engineering\\
  City University of Hong Kong\\
  Kowloon, Hong Kong SAR, China\\
  \texttt{cpeng29-c@my.cityu.edu.hk} \\
  \And
 Xubo Tang \\
  Dept. of Electrical Engineering\\
  City University of Hong Kong\\
  Kowloon, Hong Kong SAR, China\\
  \texttt{xubotang2-c@my.cityu.edu.hk} \\
  \And
 Yanni Sun \\
  Dept. of Electrical Engineering\\
  City University of Hong Kong\\
  Kowloon, Hong Kong SAR, China\\
  \texttt{yannisun@cityu.edu.hk} \\
}

\begin{document}

\maketitle
\begin{abstract}
\textbf{Motivation:} As viruses that mainly infect bacteria, phages are key players across a wide range of ecosystems.  Analyzing phage proteins is indispensable for understanding phages' functions and roles in microbiomes. High-throughput sequencing enables us to obtain phages in different microbiomes with low cost. However, compared to the fast accumulation of newly identified phages, phage protein classification remains difficult. In particular, a fundamental need is to annotate virion proteins, the structural proteins such as major tail, baseplate, etc. Although there are experimental methods for virion protein identification, they are too expensive or time-consuming, leaving a large number of proteins unclassified. Thus, there is a great demand to develop a computational method for fast and accurate phage virion protein classification.  \\
\textbf{Results:} In this work, we adapted the state-of-the-art image classification model, Vision Transformer, to conduct virion protein classification. By encoding protein sequences into unique images using chaos game representation, we can leverage Vision Transformer to learn both local and global features from sequence ``images''. Our method, PhaVIP, has two main functions: classifying PVP and non-PVP sequences and annotating the types of PVP, such as capsid and tail. We tested PhaVIP on several datasets with increasing difficulty and benchmarked it against alternative tools. The experimental results show that PhaVIP has superior performance. After validating the performance of PhaVIP, we investigated two applications that can use the output of PhaVIP: phage taxonomy classification and phage host prediction. The results showed the benefit of using classified proteins over all proteins. \\
\textbf{Availability:} The web server of PhaVIP is available via: \href{https://phage.ee.cityu.edu.hk/phavip}{https://phage.ee.cityu.edu.hk/phavip}. The source code of PhaVIP is available via: \href{https://github.com/KennthShang/PhaVIP}{https://github.com/KennthShang/PhaVIP}. \\
\textbf{Contact:} \href{yannisun@cityu.edu.hk}{yannisun@cityu.edu.hk}\\
\end{abstract}

\section{Introduction}
\label{sec:intro}
Bacteriophages, or phages for short, are viruses that can infect bacteria. They are the most widely distributed and abundant biological entities in the biosphere \cite{cobian2016viruses}, with an estimated population of more than $10^{31}$ particles \cite{lyon2017phage}. Phages play an important role in modulating microbial system dynamics by lysing bacteria and mediating the horizontal transfer of genetic material \cite{fernandez2018phage}. In addition, there are accumulating studies showing that phages have an important impact on multiple applications, such as the food industry \cite{brussow2001comparative}, disease diagnostics \cite{wang2004epitope}, engineering bacterial genomes \cite{menouni2015bacterial}, and phage therapy \cite{azimi2019phage}.

A fundamental step to promote phages' applications in these fields is phage genome annotation. Phages' proteins are highly diverse, and their current annotations are far from complete. For example, only 33\% of proteins in the RefSeq phage protein database have annotations. The annotated phage proteins can be roughly divided into two groups: virion and non-virion proteins. Phage virion proteins (PVPs) are phage structural proteins that make up phage outer protein shells \cite{kabir2022large}. They were regarded as one major evidence in phage taxonomy classification by the International Committee on Taxonomy of Viruses (ICTV). During the infection, PVP binds to the host's receptors, aiding the insertion of the phage's genetic materials into the host cell. Identifying PVPs is a fundamental step to understanding their biological properties and mechanisms of host cell binding. Due to their ubiquity and functional importance, PVPs have been leveraged in multiple downstream applications. For example, PVPs can be used as marker genes in phage host prediction \cite{boeckaerts2021predicting} and prophage identification within bacterial genomes \cite{roux2015virsorter}. Although PVPs have become commonly used features in several phage analysis tasks, accumulating research on the non-PVPs show that we may underestimate their importance. For example, non-PVPs usually play key roles in phages' lifecycles, including replication and packaging. Among non-PVPs, ``integrase'' and ``excisionase'' are two widely accepted marker genes for classifying the lifestyle of phages. Based on these marker genes, several phage lifestyle prediction methods were developed \cite{emerson2012dynamic, hockenberry2021bacphlip}. In addition, some non-PVPs are important for the binding of phage tail fibers to host receptor proteins. For example, the endoglycosidase of \textit{Salmonella virus P22} will hydrolyze lipopolysaccharide and destroy the O-specific chain for phage attachments \cite{steinbacher1996crystal}. What's more, understanding the non-PVPs can help utilize phages for engineering bacterial genomes \cite{menouni2015bacterial}, regulating gene expression, and introducing novel functions to change cell physiology \cite{feiner2015new, howard2017lysogeny}.

Because PVPs and non-PVPs have different functions, distinguishing them can extend our knowledge about phage properties and functions. Although there are experimental methods for PVP annotation, such as protein arrays and mass spectrometry, they are usually time-consuming, labor-intensive, and costly. Thus, they cannot catch up with the speed of newly identified phages by high-throughput sequencing. For example, as reported in \cite{sinha2020characterization}, only 11\% of proteins can be annotated using the mass spectrometry method. Thus, computational PVP classification is still the major choice for handling large-scale input data. The major challenge for computational PVP classification is the high diversity of proteins in phages. For example, most structural proteins encoded by tailed phages, except for portal proteins, can not be identified through pair-wise sequence alignments. According to the latest RefSeq database downloaded before Dec. 2022, 66\% of proteins are marked as ``hypothetical protein'', meaning that these proteins cannot be aligned to annotated proteins. Thus, fast and accurate computational methods to predict and classify diverged PVPs are urgently needed.

\begin{table*}[!h]
\centering
{ 
\begin{tabular}{p{3cm}p{4cm}p{4cm}cp{2.5cm}}
\hline
\textbf{Group}                                                   & \multicolumn{1}{c}{\textbf{Tools}}         & \multicolumn{1}{c}{\textbf{Encoding method}}                                  & \multicolumn{1}{c}{\textbf{Model}} \\ \hline \rowcolor{gray!10}
Traditional Machine-learning                  
                                                                 & PVPred-SCM \cite{charoenkwan2020pvpred}& $K$-mer frequency& SCM \\
                                             \rowcolor{gray!10}  & Pred-BVP-Unb \cite{arif2020pred} & Physicochemical properties & SVM \\     
                                             \rowcolor{gray!10}  & \cite{ru2019identification} & Skip-gram & RF \\
                                             \rowcolor{gray!10}  & \cite{tan2018identifying} & $K$-mer frequency (g-gap) & SVM \\
                                             \rowcolor{gray!10}  & PhagePred \cite{pan2018identification} & $K$-mer frequency (g-gap) & NB \\
                                             \rowcolor{gray!10}  & PVP-SVM \cite{manavalan2018pvp} & $K$-mer frequency, physicochemical properties & SVM \\
                                             \rowcolor{gray!10}  & PVPred \cite{ding2014identification} & $K$-mer frequency (g-gap) & SVM \\ 
                                             \rowcolor{gray!10}  & \cite{feng2013naive} & $K$-mer frequency & NB \\ \hline
\multicolumn{1}{l}{Ensemble method}   & iPVP-MCV \cite{han2021ipvp} & Position-specific scoring matrix & SVM-based ensemble model \\
\multicolumn{1}{l}{}                                            & Meta-iPVP \cite{charoenkwan2020meta}      & Probabilistic   martix & SVM, RF, NB, ANN                                   \\
\multicolumn{1}{l}{}                                            &  \cite{zhang2015ensemble} & Position-specific scoring matrix and physicochemical properties & RF-based ensemble model                                \\ \hline 
\rowcolor{gray!10}
\multicolumn{1}{l}{Deep learning} & DeePVP \cite{fang2022deepvp} & One-hot & CNN                                     \\
\rowcolor{gray!10} & VirionFinder \cite{fang2021virionfinder} & One-hot and physicochemical properties & CNN                                    \\
\rowcolor{gray!10} &      PhANNs \cite{cantu2020phanns} & $K$-mer frequency & ANN                               \\
\rowcolor{gray!10} \multicolumn{1}{l}{}  &   iVIREONS \cite{seguritan2012artificial} & Single amino acid & ANN                                 \\ \hline
\end{tabular}
}
\caption{Summary of the existing PVP classification tools.}
\label{tab:tools}
\end{table*}

\subsection{Related work}
\label{sec:relate}

To overcome the challenge of high sequence diversity, machine learning models are commonly used for classifying PVP and non-PVP. 
Most of these tools have been discussed and evaluated by several comprehensive reviews in the past three years \cite{meng2020review, nami2021application, kabir2022large}. Table \ref{tab:tools} summarizes these tools together with their employed feature encoding and machine learning algorithms.

As indicated in Table \ref{tab:tools}, four learning models (SVM, NB, RF, and SCM) are commonly used in traditional machine learning-based methods.
Ensemble-based methods utilize multiple models or training sets. For example, Meta-iPVP \cite{charoenkwan2020meta} utilizes a novel feature-representing scheme and four machine-learning algorithms to encode seven input features into a probabilistic matrix. Then, the generated probabilistic matrix is fed into the SVM model to classify PVPs. More recently, deep learning-based methods such as VirionFinder \cite{fang2021virionfinder} and DeePVP \cite{fang2022deepvp}) have been proposed for structural protein identification. Both of them use convolutional neural networks (CNNs) as classifiers. The comparison between the existing tools showed that CNN is an effective method for extracting abstract features from biological sequences \cite{kabir2022large}.

Although these tools have achieved promising performance, they still have a couple of limitations. First, except for PhANNs \cite{cantu2020phanns} and DeePVP \cite{fang2022deepvp}, all these tools are binary identifiers, which can only classify the input proteins as PVP or non-PVP. However, a more detailed multi-class classification of PVPs is also in demand to assign proteins to well-defined annotations (i.e., major tail, minor tail, and baseplate). But the best F1-score of PhANNs and DeePVP on multi-class classification can only reach nearly 0.7 on the benchmark dataset. Second, the database of the existing tools is mostly out-of-date. However, only PhANNs provided scripts for re-training or re-constructing the models as reported in \cite{kabir2022large}. Lacking this function hinders many tools from achieving more generalized and robust predictions for newly discovered phages. Third, Although one-hot encoding and $k$-mer frequency encoding are widely used in the PVP classification task, they both have disadvantages. For example, as indicated in \cite{ren2022prediction}, using one-hot encoding for protein sequences will return sparse matrices, leading to the curse of dimensionality problem in the machine learning model. $k$-mer frequency encoding fails to maintain the original amino acids' organization in the raw sequences.

\begin{figure*}[h!]
    \centering
    \includegraphics[width=0.95\linewidth]{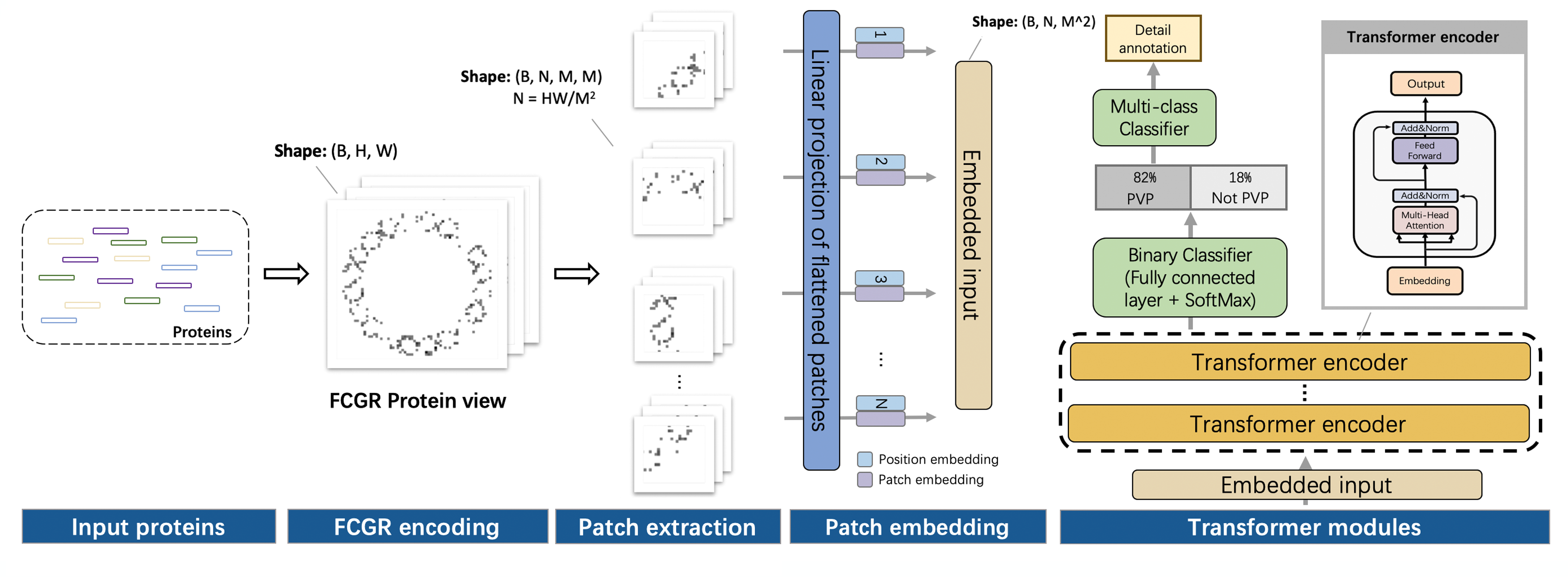}
    \caption{The pipelines of PhaVIP, which consists of three major stages: FCGR protein encoding, patch embedding, and Transformer modules. When taking a test/query protein as input, PhaVIP first classifies it into PVP and non-PVP. Only the predicted PVP will be classified into more detailed annotations.}
    \label{fig:model}
\end{figure*}

\subsection{Overview}
In this work, we present a method named PhaVIP (Phage VIrion Protein) for phage protein annotation. It has two functions. First, it can classify a protein into either PVPs or non-PVPs (binary classification task). Second, it can assign a more detailed annotation for predicted PVPs, such as major capsid, major tail, and portal (multi-class classification task with seven types of PVPs). To construct a complete and comprehensive dataset, we downloaded the latest annotations of phage proteins from the RefSeq database (Dec. 2022) to train and test PhaVIP. The pipeline of PhaVIP is shown in Fig. \ref{fig:model}. First, to address the shortages of the existing encoding methods, we employ chaos game representation (CGR) to encode proteins into images. Previous works show that using $k$-mer frequency helps distinguish proteins of different functions. However, existing models such as CNN are not optimized for learning the associations of $k$-mers and their frequencies. In our design, CGR can encode k-mer frequency into images, allowing us to leverage an image classification model, Vision Transformer (ViT), from computer vision to capture and learn the patterns from CGR images. We leverage the self-attention mechanism in ViT to learn the importance of different sub-images and their associations for protein classification. In addition, because the length of the proteins varies from $10^2$ to $10^3$, applying CGR allows encoding sequences of highly different lengths into images with the same resolution. Thus, we expect that this combination can lead to better results than existing deep learning models because of the success of the ViT in image classification. In the experiments, we tested PhaVIP on multiple datasets with increasing difficulty. The comprehensive comparison with the existing methods shows that PhaVIP renders better and more robust performance. In addition, we designed two case studies to show the application of PVPs and non-PVPs for downstream phage analysis. These case studies reveal that PhaVIP can provide useful features to improve the accuracy of phage taxonomy classification and host prediction.

\section{Methods and materials}

To use machine learning methods for classifying PVP and non-PVP, the input proteins need to be encoded into numerical values. Thus, a practical and informative sequence encoding method is crucial for classification. In this work, we applied Chaos Game Representation (CGR) to encode protein sequences. CGR is a generalized Markov chain and allows one-to-one mapping between the image and the sequence \cite{lochel2021chaos}. In addition, CGR has already shown promising results in encoding biological sequences, such as generating evolutionary trees \cite{hoang2016numerical} and finding antimicrobial resistant gene \cite{ren2022prediction}. 

Because CGR can represent protein sequences using unique images, inspired by pattern recognition problems in computer vision (CV), we apply the ViT model to extract and learn features from the CGR image. The attention mechanism in ViT can reveal the representative regions in the image and learn the associations between different parts of the image \cite{ghiasi2022vision}. Several large-scale benchmark datasets in CV have shown that ViT  outperforms traditional models, such as CNN, on image classification. All these features prompt us to employ ViT for PVP classification.

In the following sections, we will first introduce how CGR encodes protein sequences into unique images. Then, we will describe the ViT model optimized for the PVP classification task. Finally, we will introduce how we collect and generate the PVP datasets used in the experiments.

\subsection{CGR encoding}
\label{sec:cgr}
\begin{figure}[h!]
    \centering
    \includegraphics[width=0.65\linewidth]{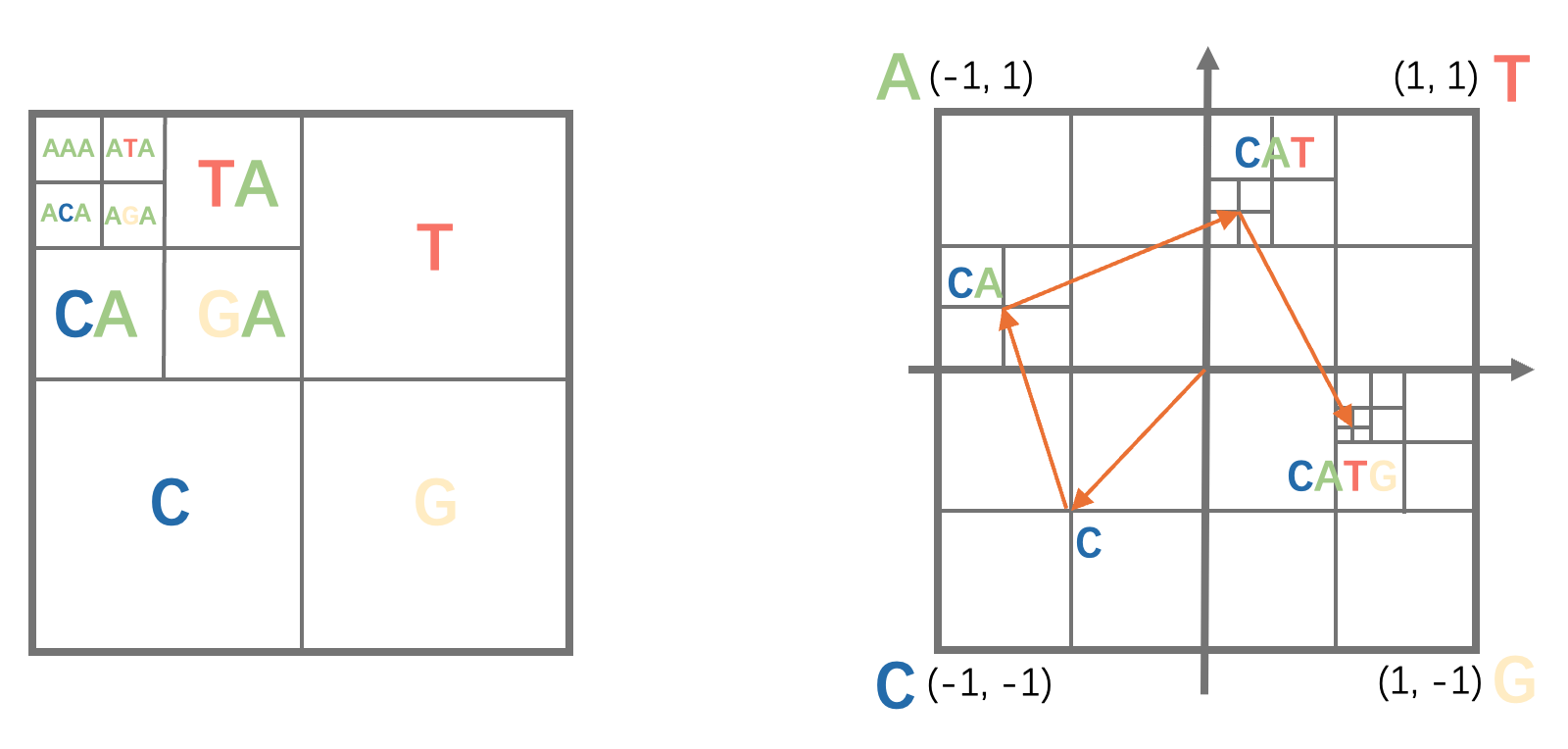}
    \caption{Applying CGR of to a toy sequence: CATG. Left: Division of the CGR space in the iterative process. (reproduced from \cite{jeffrey1990chaos}); Right: the process of determining the four pixels for CATG using CGR.}
    \label{fig:CGR_example}    
\end{figure}

The CGR was first developed to construct fractals from random inputs and later extended to encode DNA sequences \cite{jeffrey1990chaos}. The inputs to the CGR are sequences, and the outputs are numerical matrices/images representing the sequences. The basic idea of CGR is to map each nucleotide or amino acid to a unique coordinate in a 2D space. A toy example is given in the right panel of Fig. \ref{fig:CGR_example}. 

To encode the protein sequences into CGR images, we apply the n-flakes method \cite{fiser1994chaos} and use frequency chaos game representation (FCGR) to produce images of the same resolution. The equations of the n-flakes method are given in Eqn. \ref{eq:protein_cgr} \cite{fiser1994chaos}.

\begin{equation}
\label{eq:protein_cgr}
\left\{\begin{matrix}
V^x_j =& sin(\frac{2\pi j}{n}) \\
V^y_j =& cos(\frac{2\pi j}{n})
\end{matrix}\right.
\end{equation}

\noindent $j$ is the vertices ranging from 0 to n, which is set to 20 for amino acids. Then, FCGR can be generated by counting the points of the CGR based on a pre-defined grid. Specifically, the algorithm will split the CGR image into $N \times N$ regions. Then, the number of points in the region will be used as the region's frequency to compress the CGR, leading to an FCGR matrix of dimension $N \times N$ for all input sequences of different lengths. In this work, we employ the R package `Kaos' to encode protein sequences into FCGR images. Then, we set $N=64$ to generate $\mathbb{R}^{64 \times 64}$ images as the representation of the protein sequences.

\begin{figure}[h!]
    \centering
    \includegraphics[width=0.65\linewidth]{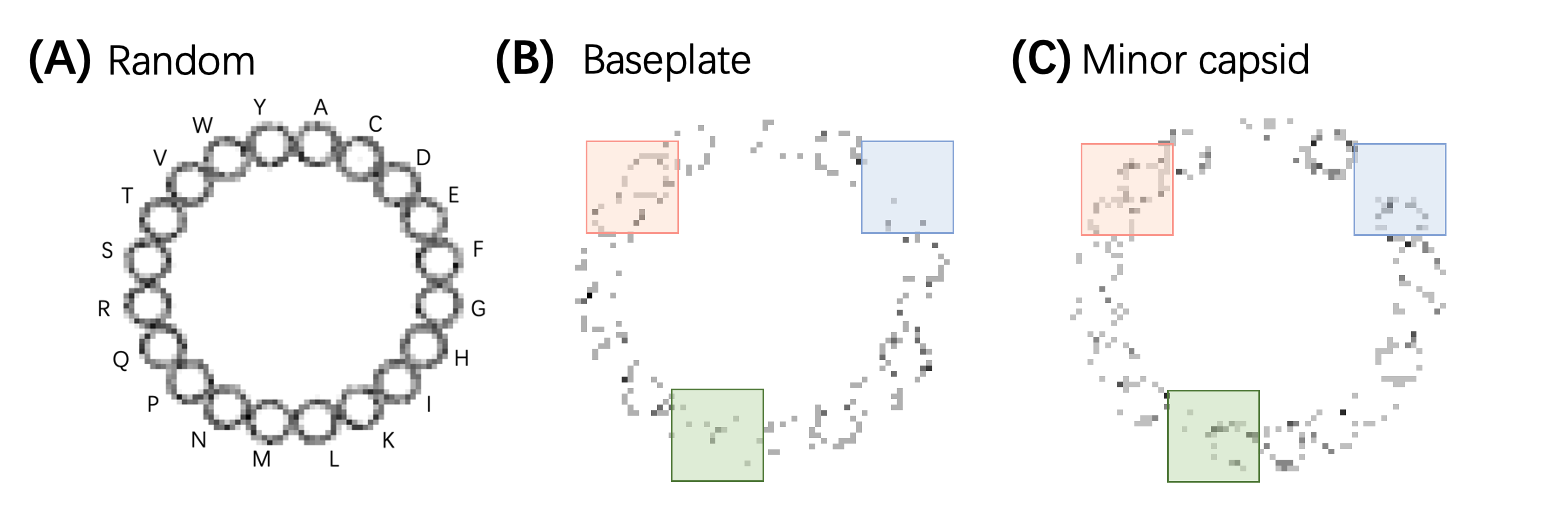}
    \caption{FCGR images for three sequences. (A): a random sequence. The order of vertices/amino acids is shown around the ring. (B): baseplate protein with RefSeq accession: \textit{YP\_009788086.1}. (C): minor capsid protein with RefSeq accession: \textit{YP\_009900655.1}. The green boxes and blue boxes in (B) and (C) show different patterns and the red boxes show exhibit patterns.}
    \label{fig:CGR_protein}    
\end{figure}

Fig. \ref{fig:CGR_protein} shows FCGR images of two different phage proteins and a random amino acid sequence. The random sequence in Panel (A) is generated by randomly choosing an amino acid for 1000 times using a uniform distribution. In contrast to the random sequence, the FCGR of baseplate protein and minor capsid protein (Fig. \ref{fig:CGR_protein} (B) and (C)) reveal more unique patterns. For example, the red patches in Fig. \ref{fig:CGR_protein} (B) and (C) exhibit a similar pattern while the blue and red patches show highly different patterns, which may signal key sequence features that can distinguish the baseplate and minor capsid proteins.  The patches indicate the distribution of short motifs ending with different amino acids. These patches and their relationships/associations with other patches can be learned by our ViT model to improve classification accuracy.

\subsection{Basic structure of ViT}
After encoding the protein sequences into $\mathbb{R}^{64 \times 64}$ images, we employ ViT for PVP classification. As shown in Fig. \ref{fig:model}, the inputs to our ViT model are FCGR images, and the output of the ViT is the probability of the protein being PVP. If the protein is predicted as PVP, our ViT will assign a more detailed annotation for the PVP.

\subsubsection{Patch splitting and embedding}
To feed an FCGR image to ViT, we will reshape the FCGR image $G \in \mathbb{R}^{64 \times 64}$ into a sequence of flattened 2D patches $g \in \mathbb{R}^{N \times M^2}$, where the dimension of each image patch is $\mathbb{R}^{M \times M}$, and $N$ is $64^2/M^2$. In our design, $M$ is set to 16 by default, and the length of the input sequence $N$ will be 16. Then, we can use Eqn. \ref{eq:embed} to generate inputs to the Transformer model. 

\begin{equation}
\label{eq:embed}
\left\{\begin{matrix}
E_t =& [g_m^1H_e; g_m^2H_e;...; g_m^NH_e] \\
E_m =& IH_m \\
Z^{(0)} =& E_m + E_t
\end{matrix}\right.
\end{equation}

\noindent Here, $g_m^i \in \mathbb{R}^{1 \times M^2}$ is the flattened 2D patch at position $i$, corresponding to a ``word'' token in Transformer for natural language modeling \cite{vaswani2017attention}. $I$ is the index of the position of each patch $x_m^i$ in the input FCGR image. $H_e$ and $H_m$ are learnable linear projection matrices for image patch and positional embedding, respectively.

\subsubsection{The Transformer model}

The architecture of the Transformer model in Fig. \ref{fig:model} is the same as the original design in \cite{vaswani2017attention}. The equations of the Transformer are listed in Eqn. \ref{eq:trans}. The first function is the multi-head attention mechanism (\textit{MSA} layer), which can extract the importance of patches and learn their associations.  Then linear projections (\textit{MLP} layer) are employed to capture information from each patch simultaneously. Layer normalization (LN) \cite{wang2019learning}, and residual connections \cite{baevski2018adaptive} are applied before and after each block to prevent gradient exploding and gradient vanishing, respectively. In the last layer, we use the SoftMax function to estimate the probability of a protein being a PVP. If the protein is predicted as a PVP, $Z^{(2)}$ (Eqn. \ref{eq:trans}) will be fed to a multi-class classifier to predict a more detailed annotation.

\begin{equation}
\label{eq:trans}
\left\{\begin{matrix}
Z^{(1)} &=& MSA(Z^{(0)}) + Z^{(0)} \\
Z^{(2)} &=& MLP(LN(Z^{(1)})) + Z^{(1)} \\
y^{binary} &=& SoftMax(LN(Z^{(2)})) \\
y^{multi-class} &=& SoftMax(LN(Z^{(2)})) \\
\end{matrix}\right.
\end{equation}

\subsubsection{Model training}
Because there are two tasks in PhaVIP: classifying the PVP and non-PVP sequences (binary classification task) and classifying seven types of PVP (multi-class classification task), we train classifiers for them separately. As introduced in \cite{devlin2018bert}, pre-training the Transformer model can improve the performance of the downstream task. Thus, we first apply an end-to-end method to train the binary classification model. Then, we fix the parameters in the Transformer encoder and fine-tune a new classifier layer for the multi-class classification model. Binary cross-entropy (BCE) loss and L2 loss are employed for the binary classification and multi-class classification, respectively. We employ Adam optimizer with a learning rate of 0.001 to update the parameters for both tasks. The models are trained on HPC with the GTX 3080 GPU unit to reduce the running time.

\subsection{Data collection and experimental setup}
\label{sec:data}
Although several PVP datasets have been constructed \cite{kabir2022large}, the latest dataset constructed by \cite{cantu2020phanns} was based on the protein annotations released before June 2020. In addition, some annotations of phage protein can be updated regularly in the RefSeq database. For example, as the author of DeePVP \cite{fang2022deepvp} reported, the protein \textit{YP\_006383517.1} was not annotated as PVP until Oct 2021, and this protein was re-annotated as a tail protein in the current version. Thus, in this work, we updated the PVP classification dataset by downloading all the latest annotations from the RefSeq viral protein database (Dec 2022). Following the guidelines of the third-party review \cite{kabir2022large}, we first recruited proteins that belong to phages. Then, the proteins with low-confidence annotations, such as ``hypothetical protein'', ``similar to'', ``xx-like'', ``unnamed'', and ``putative'' were removed. We extracted structural protein sequences by searching the keywords, such as ``portal'', ``capsid'', ``tail'', ``fiber'', ``tape measure'',  ``baseplate'', and ``structural''. The non-structural proteins were searched using the enzymes' names, such as annotations ending with ``ase''. In addition, we also used other keywords, such as ``transcription'', ``holin'', ``lysin'', and ``regulator'', to construct the non-PVP set. To remove the potential redundant sequences, we employed CD-HIT \cite{li2006cd} to cluster sequences that have above 90\% similarity and used the longest sequence to represent each cluster. Finally, our dataset contains 35,213 PVP sequences and 46,883 non-PVP sequences.

\subsubsection{Splitting the dataset}
We split our PVP dataset with increasing difficulty when constructing the training and test set. There are two tasks for PVP classification: classifying the PVP and non-PVP sequences (binary classification) and predicting the PVP types (multi-class classification). In the binary classification task, we use all the proteins for the data partition. In the multi-class classification task, we use the protein annotated with ``portal'', ``major capsid'', ``minor capsid'', ``major tail'', ``minor tail'', ``baseplate'', and ``tail\_fiber'' to construct the multi-class classification dataset. All the remained PVPs are labeled as ``other''. We select the seven classes because they represent the dominant structural protein roles and contain enough sequences ($>$ 100) for training and testing. 

\paragraph{Splitting by time}
As mentioned in \cite{kabir2022large}, splitting training and test set by time is a widely used data partition method, which mimics the application scenario of using known PVPs to discover new ones. In this dataset, proteins released before Dec. 2020 comprise the training set, while proteins released after that comprise the test set. Finally, we have 27,704 PVP sequences and 36,778 non-PVP sequences for training, and 7,509 PVP sequences and 10,103 non-PVP sequences for testing in the PVP classification task. To balance the dataset, we randomly sampled non-PVP sequences to maintain the same number of samples in the binary classification as suggested in \cite{kabir2022large}. In the multi-class classification, we keep the original data distribution following \cite{cantu2020phanns}.

\paragraph{Splitting by similarity}
To test PhaVIP's performance in classifying diverged PVP, we constructed a hard case where the test sequences share low similarity with the training proteins. Specifically, we applied the all-against-all BLASTP search to our PVP dataset and calculated the product of the pairwise identity and alignment coverage, which is the ratio of aligned length to the length of the query sequence. Then, we employed the data partition strategy proposed in \cite{petti2022constructing} to create training and test data with a specified maximum similarity between train and test. In this work, we chose 0.4, 0.5, 0.6, 0.7, 0.8, and 0.9 as the thresholds and employed stratified sampling to split the training and test sets. 

\subsubsection{Metrics}

As mentioned in \cite{kabir2022large}, the widely used metrics for evaluating PVP classification performance are precision, recall, and F1-score. Their formulas are listed in Eqn.\ref{m1}, Eqn. \ref{m2}, and Eqn. \ref{m3}: 

\begin{equation}
    \label{m1}
    precision = \frac{ TP }{TP+FP}  
\end{equation}

\begin{equation}
    \label{m2}
    recall = \frac{ TP }{TP+FN}  
\end{equation}

\begin{equation}
    \label{m3}
    F1\raisebox{0mm}{-}score = \frac{ 2*precision*recall }{precision+recall}
\end{equation}

\noindent For binary PVP classification, true positive (\textit{TP}), false negative (\textit{FN}), and false positive (\textit{FP}) represent the number of corrected identified PVPs, the number of PVPs misclassified into non-PVPs, and the number of falsely identified PVPs, respectively. We will also report the Area Under the ROC Curve (AUCROC) for comparison. For the multi-class classification task, we will calculate precision, recall, and F1-score for each class.

\section{Result}
In the experiment, we validate our pipeline on several datasets and compare PhaVIP against the state-of-the-art methods mentioned in the third-party review \cite{kabir2022large}, including VirionFinder \cite{fang2021virionfinder}, PhANNs \cite{cantu2020phanns}, DeePVP \cite{fang2022deepvp}, Meta-iPVP \cite{charoenkwan2020meta}, PVP-SVM \cite{manavalan2018pvp}, and PVPred-SCM \cite{charoenkwan2020pvpred}. Out of these tools, only PhANNs provided source codes for re-training or updating the reference database. Thus, we are able to retrain PhaANNs for both the binary and the multi-classification tasks using the suggested hyperparameters. Other tools did not provide a retraining function. Thus, we applied them to the test data directly.

In the following sections, we will first evaluate the PVP classification performance. Then, following \cite{fang2022deepvp}, we will show a case study of classifying PVP on the \textit{mycobacteriophage} PDRPxv genome, a newly identified phage that is a candidate therapy for pathogenic \textit{Mycobacterium}. Finally, we investigate whether using classified PVPs and non-PVPs can benefit two important phage analysis tasks: phage taxonomy classification and host prediction.

\subsection{Performance on the benchmark dataset split by time}

\begin{figure}[h!]
    \centering
    \includegraphics[width=0.5\linewidth]{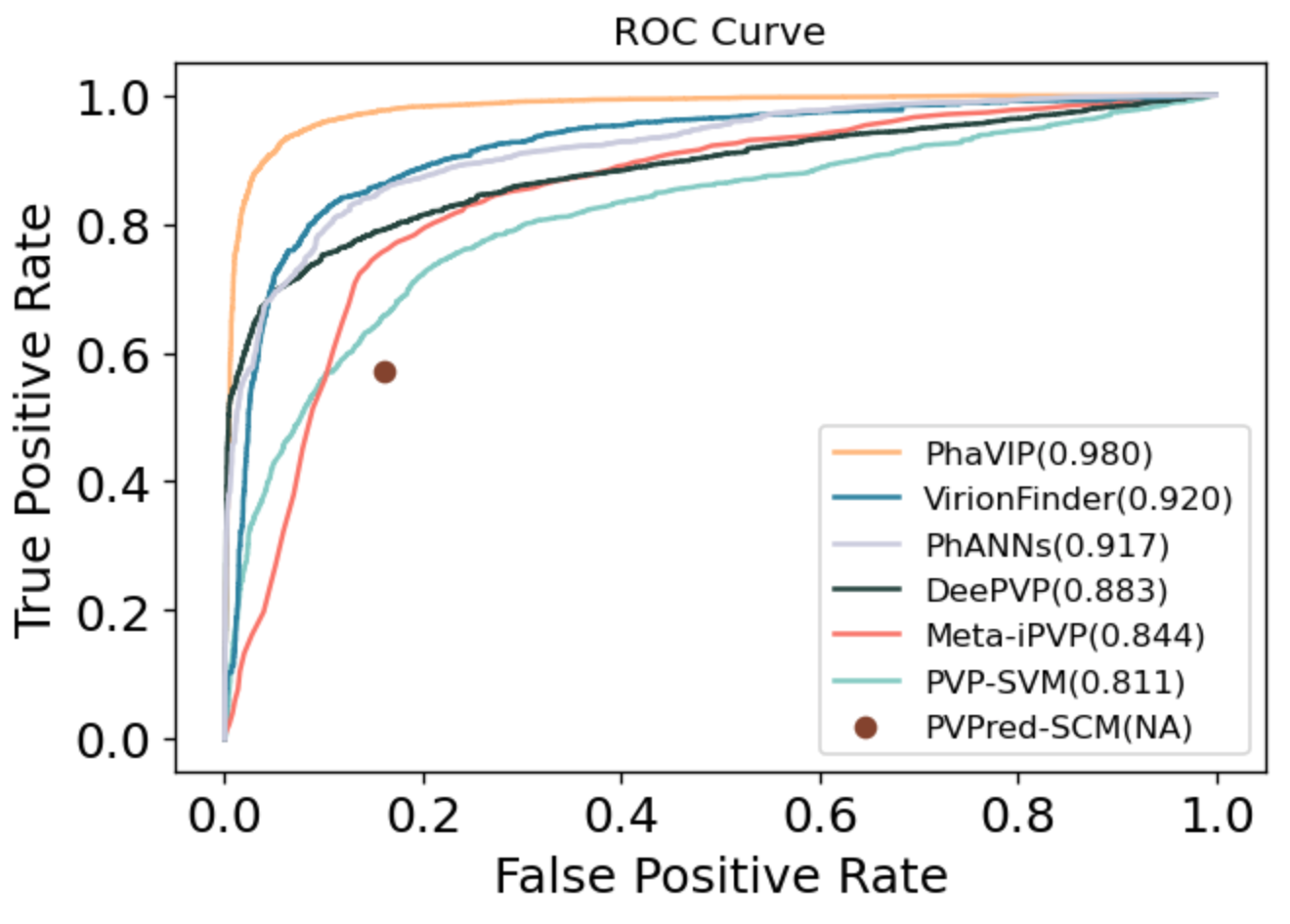}
    \caption{The ROC curves of the binary PVP classification by different tools. The number following the tool name is the value of the AUCROC. PVPred-SCM does not output a score for the prediction, and thus only TPR and FPR are reported.}
    \label{fig:roc}    
\end{figure}

To improve the robustness of the model, we trained PhaVIP and PhANNs using ten-fold cross-validation. First, we split our training set into ten subsets. Then, we iteratively selected nine subsets for training and one subset for validation. The model that achieves the best performance on the validation set was kept for future experiments. For other methods, we used the provided models with the suggested parameters on the test proteins. The ROC curves of all the methods are shown in Fig. \ref{fig:roc}. The AUCROC reveals that PhaVIP has more reliable results on the dataset split by time. Because PVPred-SCM does not output a score of the prediction, we only report its recall and false positive rate.

In order to show the classification performance in real application scenarios, we also recorded the precision, recall, and F1-score of all tested tools under their default score cutoffs in Fig. \ref{fig:bytimebinary} and Table S1 in the supplementary file. The results reveal that PhaVIP and DeePVP achieve the highest precision (0.94). Meanwhile, PhaVIP has a higher recall than DeePVP.

\begin{figure}[h!]
    \centering
    \includegraphics[width=0.5\linewidth]{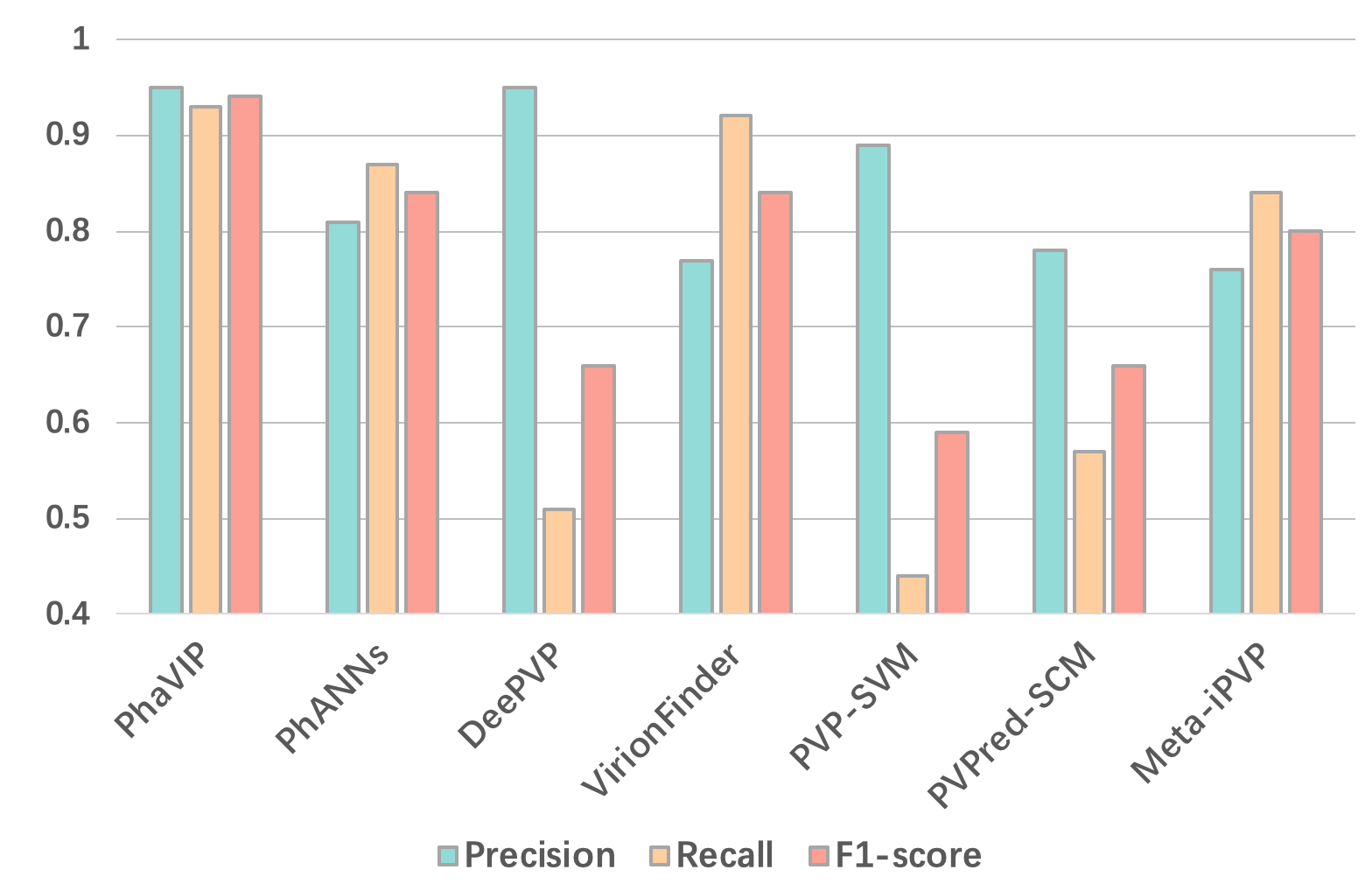}
    \caption{The classification performance of the binary PVP classification under the default/suggested thresholds.}
    \label{fig:bytimebinary}    
\end{figure}

Next, we examine the performance of PhaVIP in the multi-class classification task. Of the available tools, only PhANNs and DeePVP enable more detailed annotation of PVPs. However, classes/labels in the original design of PhANNs and DeePVP are different from ours and only PhANNs allows us to retrain on the multi-classification dataset. Thus, we retrained PhANNs and compared it with PhaVIP. The F1-score of each class is shown in Fig. \ref{fig:bytimemulti}, and the detailed confusion matrix can be found in Table S2 in the supplementary file. The results clearly show that the multi-class classification task is harder than classifying the PVP and non-PVP. The possible reasons are the smaller training sets and highly unbalanced classes. Although both PhaVIP and PhANNs used the weighted loss method to balance the training classes, the performance of the small class (minor capsid) is still unsatisfactory. Nevertheless, PhaVIP can achieve better performance in all the classes, especially in the small ones.

\begin{figure}[h!]
    \centering
    \includegraphics[width=0.5\linewidth]{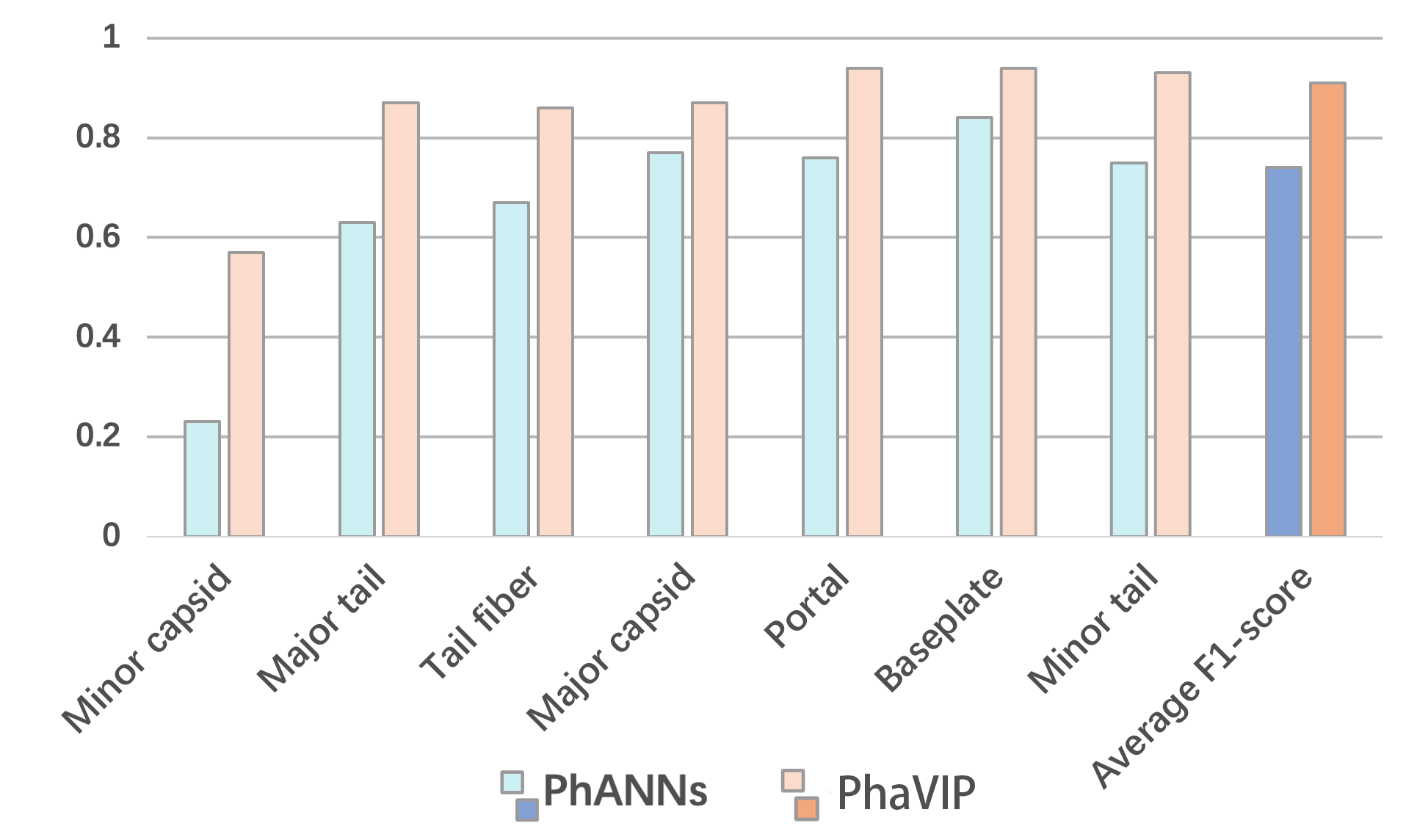}
    \caption{The performance of the multi-class classification. X-axis: the name of each PVP class. The order of the names is ranked by the class size. Y-axis: F1-score.}
    \label{fig:bytimemulti}    
\end{figure}

\subsection{Performance on the low-similarity dataset}
\begin{figure}[h!]
    \centering
    \includegraphics[width=0.5\linewidth]{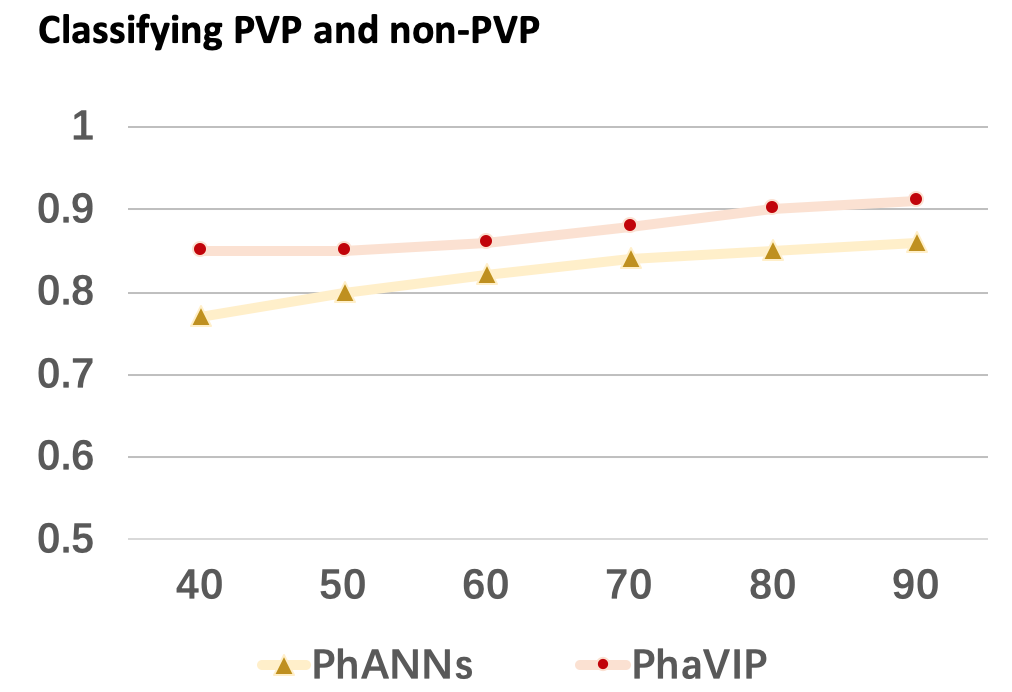}
    \caption{The binary classification performance on the low similarity dataset. X-axis: The maximum value of $identity \times coverage$ between train and test sets. . Y-axis: F1-score.}
    \label{fig:bysimilarityb}    
\end{figure}

\begin{figure}[h!]
    \centering
    \includegraphics[width=0.5\linewidth]{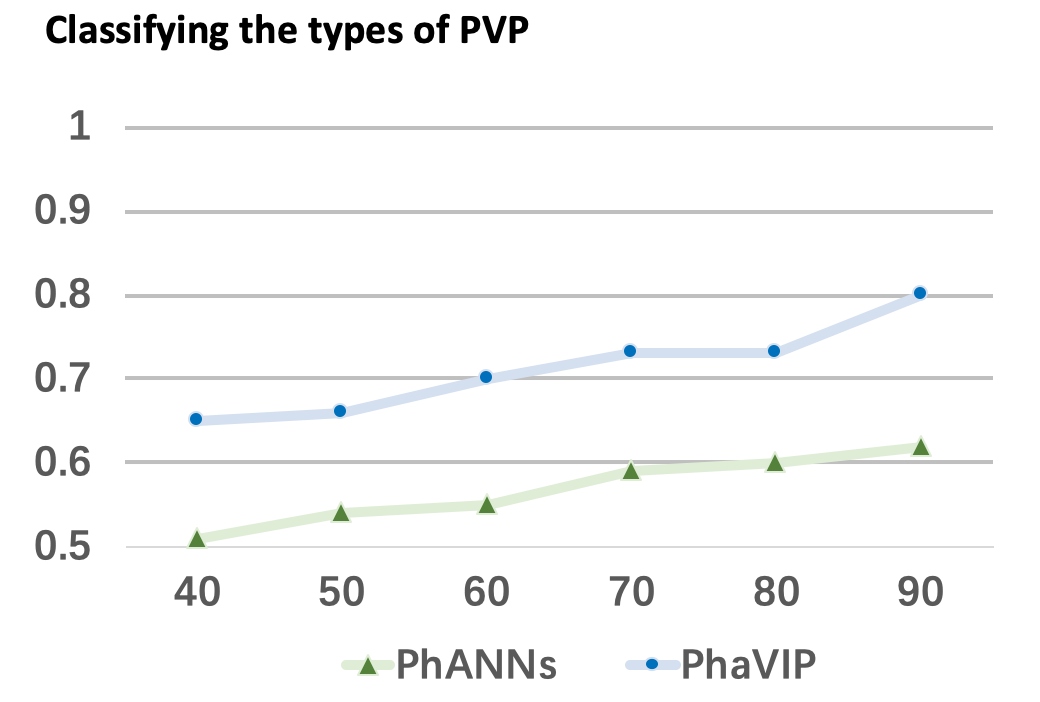}
    \caption{The multi-class classification performance on the low similarity dataset. X-axis: The maximum value of $identity \times coverage$. Y-axis: F1-score.}
    \label{fig:bysimilaritym}    
\end{figure}

It is usually much harder to annotate diverged proteins. As mentioned in Section \ref{sec:data}, we use the $Identity \times Coverage$ as the similarity measurement and control the maximum similarity between the training and test set. We generated six datasets with decreasing similarity for the binary classification task and multi-class classification task separately. The F1-scores of PhaVIP and PhANNs are shown in Fig. \ref{fig:bysimilarityb} and Fig. \ref{fig:bysimilaritym}. The detailed confusion matrix of the classification can be found in Table S3-S14 in the supplementary file.

As expected, with the increase of the train-vs-test similarity, the F1-score of both methods increases. The gap between PhaVIP and PhANNs clearly reveals that our model competes favorably against PhANNs on a wide range of similarities in both binary and multi-class classification tasks.

\subsection{Case study: annotating proteins on the \textit{mycobacteriophage} PDRPxv genome}
In this case study, we employed PhaVIP to annotate the proteins translated from \textit{mycobacteriophage} PDRPxv, which is recently identified as a candidate therapy for  \textit{Mycobacterium}. According to  \cite{sinha2020characterization}, totally there are 107 predicted proteins in the PDRPxv genome. The authors identified 12 PVP using the mass spectrometry method and 12 non-PVP using the alignment method (BLAST). The functions of the other 83 proteins remain unknown. Because PDRPxv is not part of the RefSeq dataset, we can evaluate PhaVIP by comparing PhaVIP's predictions with the 24 annotations derived by the mass spectrometry method and BLAST. 

\begin{table}[h!]
\centering
\begin{tabular}{ccccc}
\hline
Tools    & PhaVIP & DeePVP & PhANNs & VirionFinder \\ \hline
F1-score & \textbf{0.88} & 0.85   & 0.83   & 0.64         \\ \hline
\end{tabular}
\caption{F1-score of classifying proteins in mycobacteriophage PDRPxv genome.}
\label{tab:case1}
\end{table}

\begin{figure}[h!]
    \centering
    \includegraphics[width=0.5\linewidth]{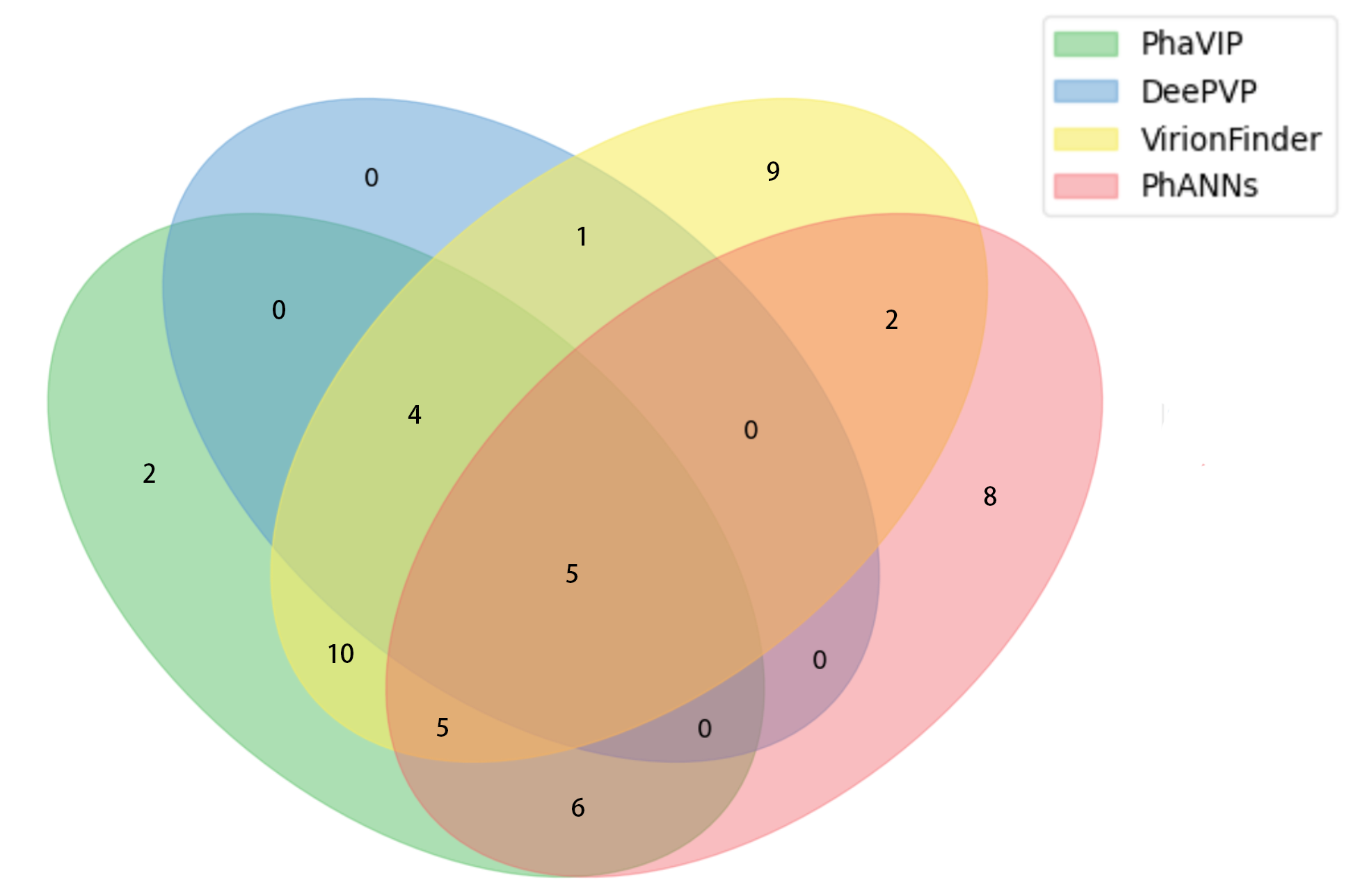}
    \caption{The Venn diagram of the complete PVP classification results of four best machine-leanring methods on \textit{mycobacteriophage} PDRPxv.}
    \label{fig:case1}  
\end{figure}

We used the 24 annotated proteins as input and tested the performance of the best four tools (in the benchmark experiment in Fig. \ref{fig:bytimebinary}). As shown in Table \ref{tab:case1}, PhaVIP has better performance than other tools. In addition, all the machine learning-based methods are able to predict the function of the remaining 83 proteins, demonstrating the utility of the learning-based method for PVP classification. We used the Venn diagram to visualize the relationship between the predicted PVP sets. As shown in Fig. \ref{fig:case1}, PhaVIP, VirionFinder, PhANNs identified more PVPs than DeePVP. This is consistent to the observation of DeePVP's low recall in Fig. \ref{fig:bytimebinary}. In addition, 93\% of PVPs predicted by PhaVIP are also classified as PVPs by other methods, which is higher than PhANNs and VirionFinder.

\subsection{Using classified proteins in two important applications}

It is widely known that phage proteins play essential roles in taxonomy classification and host prediction. In this section, we investigate the roles of PVPs and non-PVPs in these two tasks. 

\subsubsection{Phage taxonomy classification}
Recently, many new phages have been identified using high-throughput sequencing, especially metagenomic sequencing.  vConTACT 2.0 \cite{eloe2019towards} is a widely used and robust tool for phage taxonomy classification, as reported in the phage taxonomy review \cite{zhu2022phage}. It applies protein organization conservation for phage classification. Specifically, vConTACT 2.0 calculates the p-value that estimates the significance of two phage sequences sharing an observed number of proteins. Then, a protein-sharing network is constructed based on the p-value, and a clustering algorithm is applied to group ``similar'' sequences into the same cluster. Then the known labels of the reference genomes in the cluster will be passed to other sequences in the same cluster.

\begin{figure}[h!]
    \centering
    \includegraphics[width=0.65\linewidth]{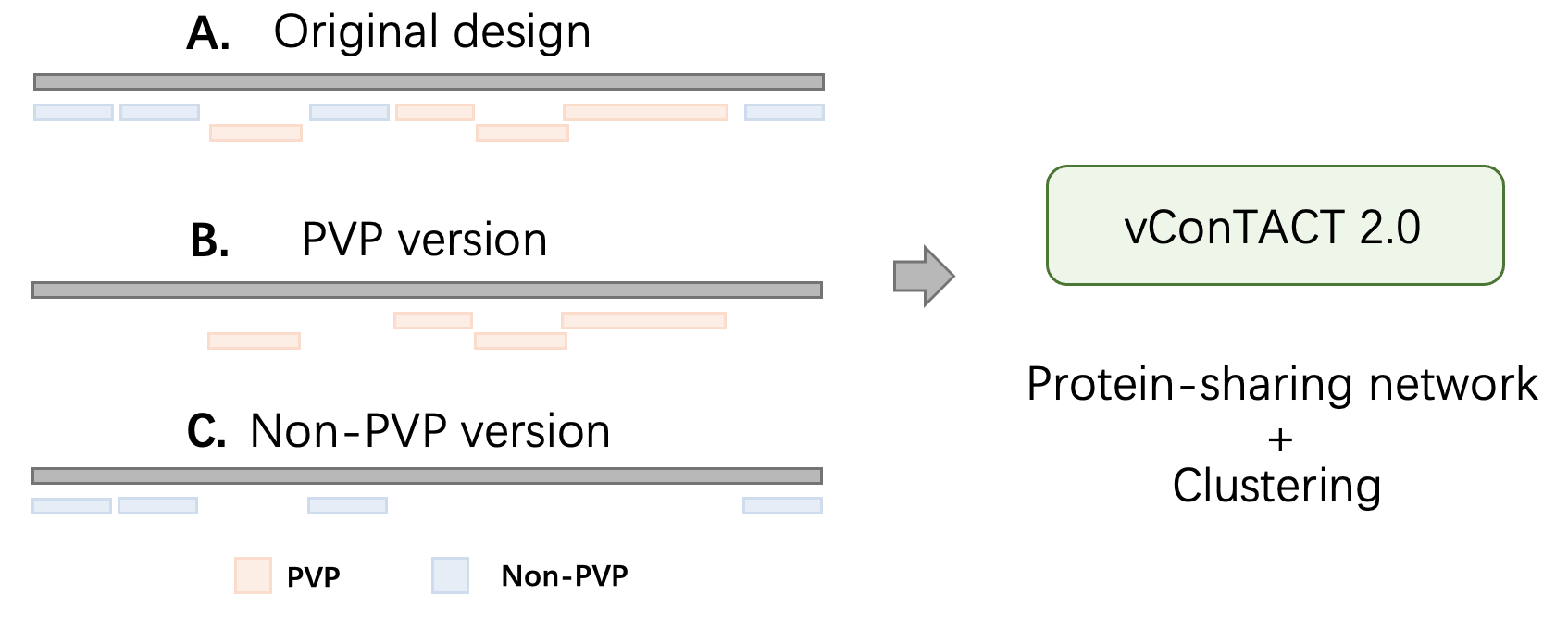}
    \caption{Three versions of vConTACT 2.0. A. The original design of vConTACT 2.0 uses all the proteins from the phage genome to construct the protein-sharing network. B. PVP version of vConTACT 2.0. C. non-PVP version of vConTACT 2.0.}
    \label{fig:vcontact}    
\end{figure}

Although vConTACT has high accuracy in classifying complete or near complete phage sequences, its running time complexity is high because of large-scale pairwise alignments. Thus, instead of using all proteins (Fig. \ref{fig:vcontact} A), we propose to only use PVPs or non-PVPs to evaluate the similarity between phages. In particular, because PVPs have successful applications in phylogenetic tree construction, we expect that using just PVPs can achieve comparable accuracy of phage classification as using all proteins. Thus, in this experiment, we use just PVP or non-PVP when running vConTACT 2.0 and evaluate how PVP or non-PVP affects the classification results. First, we downloaded the benchmark dataset provided by \cite{zhu2022phage}. This dataset was constructed using 1460 RefSeq phage sequences from the latest ICTV 2022 taxonomy. It was split by time: 80\% of the sequences in each family were used as the training set, and the remaining sequences were used as the test set. Second, we applied prodigal \cite{hyatt2010prodigal} to predict and translate proteins from the phage genomes in training and test sets. PhaVIP is then employed to annotate each protein. Finally, we used predicted PVPs and non-PVPs to predict the taxonomy via vConTACT, respectively. Fig. \ref{fig:vcontact} B and C sketched the pipelines.

\begin{figure}[h!]
    \centering
    \includegraphics[width=0.65\linewidth]{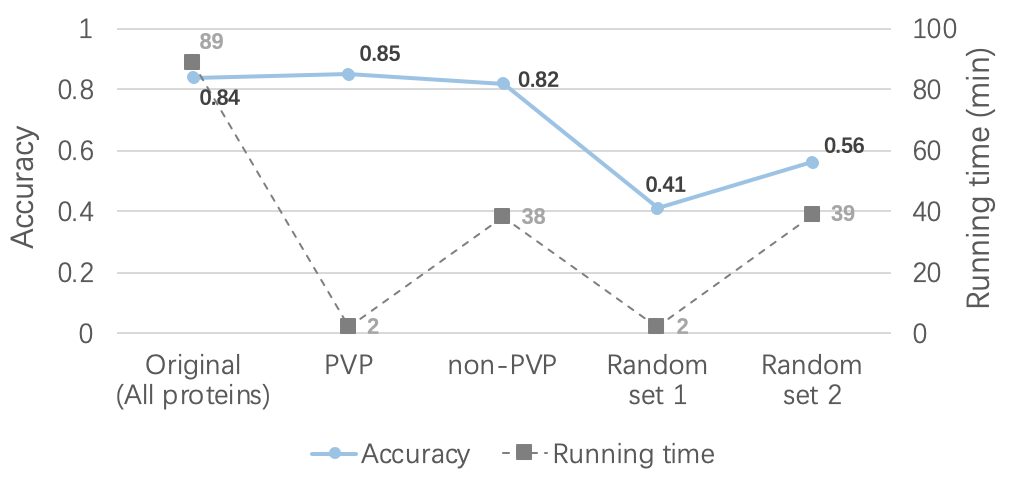}
    \caption{vConTACT taxonomy classification results using different sets of proteins. ``Random set 1'' and ``Random set 2'' represent randomly selected protein sets, which have the same number of proteins as PVP and non-PVP set, respectively.}
    \label{fig:vcontact_result}    
\end{figure}

The taxonomy classification results in Fig. \ref{fig:vcontact_result} show that  the PVP version of vConTACT 2.0, which only used PVP for taxonomy classification, can achieve almost the same performance as the regular vConTACT 2.0. In addition, because PVP only accounts for nearly $1/5$ of the total predicted proteins, using PVP for taxonomy classification can reduce the running time significantly. Because running PhaVIP only takes about seven minutes for all proteins, even with the preprocessing by PhaVIP, the total running time of taxonomy classification by vConTACT 2.0 reduces from 89 minutes to 9 minutes. Using non-PVP for taxonomy classification can also reduce the running time. But the accuracy is 3\% lower than using PVP. 

A fair question is whether using any set of randomly chosen proteins can achieve similar accuracy with reduced running time. To answer this question, we randomly chose the same number of proteins as the PVP set and non-PVP set for taxonomy classification, respectively. In this experiment, PVP and ``Random set 1'' contain 7,105 proteins, and non-PVP and ``Random set 2'' contain 29,321 proteins. The results in Fig. \ref{fig:vcontact_result} indicate that using a random set of proteins cannot achieve comparable accuracy as using just PVPs.  In addition, vConTACT 2.0's results using ``Random set 1'' is worse than ``Random set 2'' probably because the number of proteins in ``random set 2'' is larger than ''Random set 1''. Overall, these results show that PhaVIP can help select a small subset of important proteins for taxonomy classification.

\subsubsection{Phage host prediction}
The hosts of the phages are mainly bacteria. Identifying the phage-host relationship helps decipher the dynamic relationship between microbes. In addition, because of the fast rise of antibiotic-resistant pathogens, phage therapy has become a potential alternative to antibiotics for killing the ``superbugs'' \cite{lee2020osong}. Thus, predicting the phage host is important to both fundamental research and phages' applications. 

\begin{figure}[h!]
    \centering
    \includegraphics[width=0.65\linewidth]{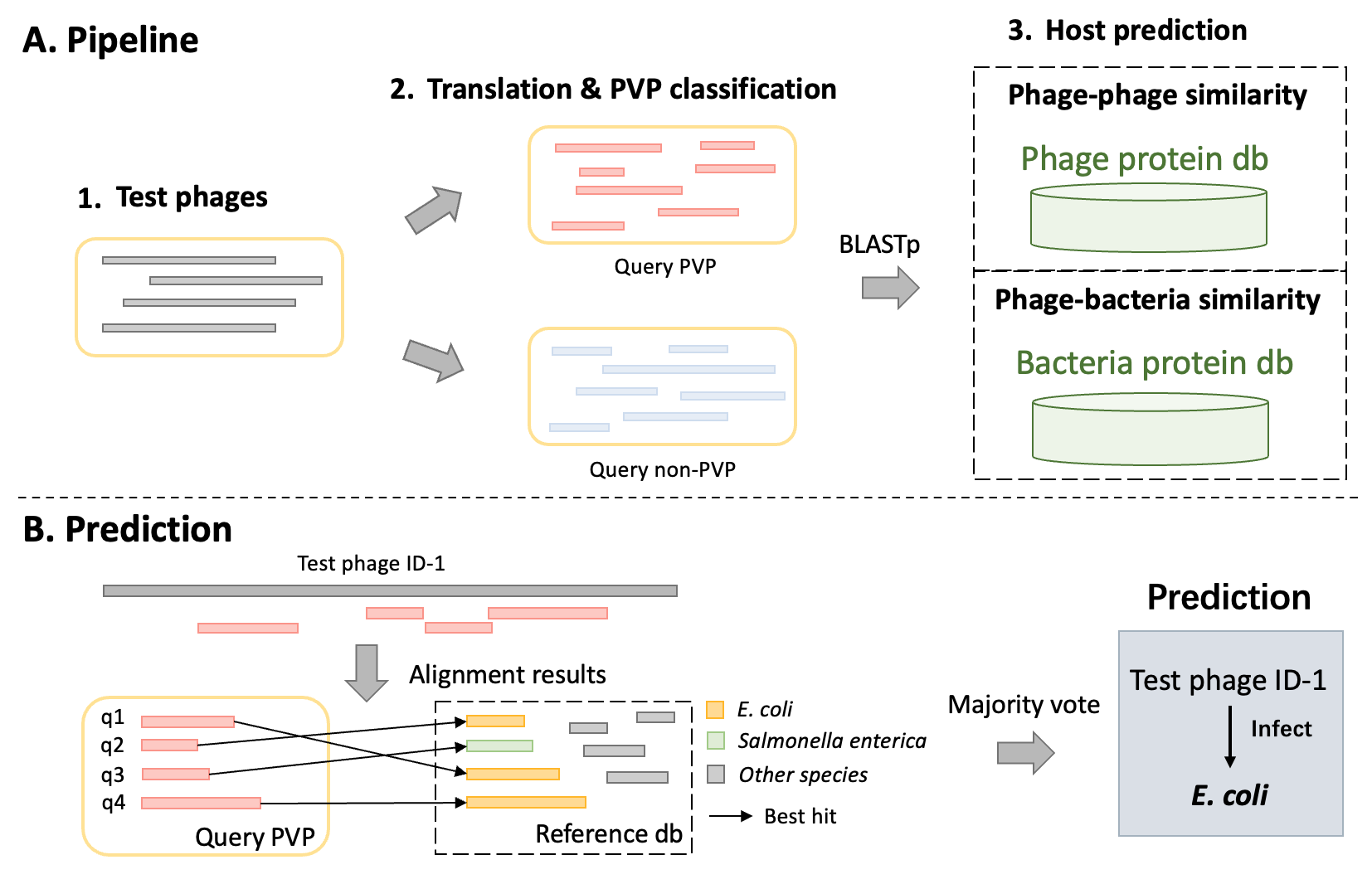}
    \caption{The pipeline of using similarity search for host prediction. A: the similarity search based host prediction. We implemented two pipelines using phage protein and bacterial protein as the reference databases, respectively. B: the majority vote method for generating the final host prediction.}
    \label{fig:host} 
\end{figure}

As reported in \cite{shang2022cherry}, sequence similarity can be utilized  for host prediction. If two phages share similar protein organizations, they tend to infect the same host. In addition, sequence similarity between phages and bacteria may help host prediction because phages can mobilize host genes \cite{howard2017lysogeny}. Thus, we developed a host prediction pipeline based on protein similarity in order to investigate how different types of proteins affect the prediction performance. The sketch of the pipeline is shown in Fig. \ref{fig:host}.

First, we downloaded the widely used benchmark dataset for host prediction \cite{shang2021predicting, shang2022cherry}. The training set contains 1,306 phage-host interactions collected in and before 2015, and the test set contains 634 interaction pairs after 2015. Every phage is unique, and some of them infect the same host. The training and test sets share 59 host species. Because the alignment-based method cannot predict new labels, we only keep 423 phages in the test set that infect these 59 species for this experiment. 

\begin{figure}[h!]
    \centering
    \includegraphics[width=0.65\linewidth]{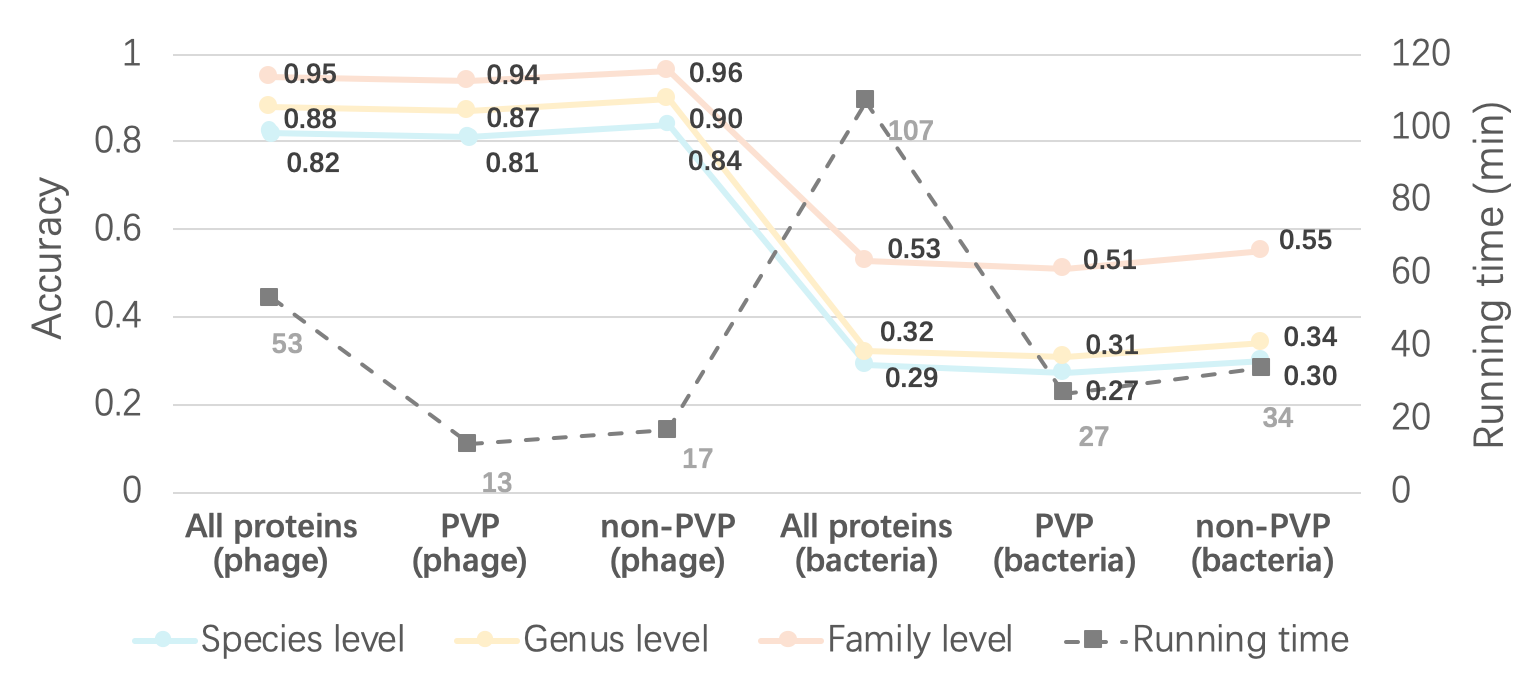}
    \caption{Host prediction results after PVP classification. ``(phage)'' and ``(bacteria)'' refer to the similarity search against the phage protein and bacterial protein databases, respectively.}
    \label{fig:host_result}    
\end{figure}

Second, we create the reference protein databases using the predicted proteins from all the phages in the training set and their hosts. As shown in Fig. 12 A, we save the proteins from phages and their hosts in two databases, respectively. Each protein has a taxonomic label. A phage protein's label is determined by its host. A bacterial protein's label is from itself. When there is a query/test phage, we predict its proteins and annotate PVP and non-PVP using PhaVIP. Then, we align the PVP proteins to the phage and bacterial protein databases and record each PVP's best alignments against two databases, respectively.  The labels of the best aligned proteins are used for host prediction. Because there are multiple proteins, we applied the majority vote as shown in Fig. \ref{fig:host} B. Specifically, the label with the most votes is assigned as the host of the phage. An example is given in Fig. \ref{fig:host} B, where three proteins were labeled as \textit{E. coli} and one protein was labeled as \textit{Salmonella enterica}. Thus, the final predicted host of this phage is \textit{E. coli}. Because we have two different databases, we record the results using the phage database and bacterial database separately. As a control experiment, we also repeated the host prediction process using only non-PVPs and all proteins. The host prediction results at different ranks from species to family are shown in Fig. \ref{fig:host_result}.

The results reveal that the similarity search against the phage protein database always has better performance than against the bacterial protein database. This phenomenon is also noted by the existing host prediction tools. As reported in \cite{shang2022cherry}, the tools based on phage-phage similarity usually have better performance than those based on phage-bacteria similarity in the experiments. In addition, we found that non-PVP can achieve better performance in host prediction tasks across different taxonomy levels and databases. 
A plausible explanation is that the host cell attachment process is complicated and involves many proteins. Some non-PVPs, such as endoglycosidase and endosialidase \cite{steinbacher1996crystal, stummeyer2005crystal}, also play key roles in the infection, and they are likely to be host-specific. Therefore, using just PVP for host prediction does not necessarily produce a better result, which does not agree with some previous conclusions \cite{boeckaerts2021predicting}. We may have underestimated the importance of non-PVPs in host prediction tasks. 

\section{Discussion}
In this work, we present a novel PVP classification tool, named PhaVIP, that combines CGR and ViT for protein encoding and PVP classification. PhaVIP has two functions: predicting the PVP and Non-PVP and predicting the type of the PVP. CGR-based encoding can convert proteins with different lengths into images with the same resolution. For each protein, it embeds the $k$-mer frequency into a unique image, allowing us to employ the state-of-the-art image classification model, ViT, to  learn the importance and associations between different parts of a CGR image. As shown in all of our experiments, ViT shows better and more robust performance in both binary classification and multi-class classification tasks. We also demonstrated that phage taxonomy classification and host prediction can benefit from using classified proteins rather than all proteins.

Although PhaVIP has greatly improved PVP classification, we have several goals to optimize or extend PhaVIP in our future work. First, although PhaVIP can render good performance in binary classification, there is still room to improve multi-class classification, especially on the low-similarity data. We will investigate whether some multi-objective loss balancing methods can be incorporated into PhaVIP to overcome the imbalance problem. Second, as mentioned in Section \ref{sec:data}, our multi-class classification includes seven common labels of PVP. There still exist some small groups of PVP, such as tail sheath and collar proteins, that do not contain enough annotated samples for training. We will explore whether we can employ the few-shot learning-based method to learn features from the classes with few labels. This can be used to provide more detailed annotations for further analyzing phages.

%
%


\section*{Funding}
City University of Hong Kong (Project 9678241 and 7005453) and the Hong Kong Innovation and Technology Commission (InnoHK Project CIMDA).

\bibliographystyle{unsrt}  
\bibliography{references}

\end{document}